\documentclass[pre,reprint,nofootinbib,superscriptaddress]{revtex4-2}
\pdfoutput=1

\usepackage{amssymb, amsmath, amsmath, amsfonts}
\usepackage[colorlinks,linkcolor=blue,urlcolor=blue,citecolor=blue]{hyperref}
\usepackage{graphics}
\usepackage{physics}
\graphicspath{{figs/}}

\usepackage[english]{babel}

\usepackage{xcolor}
\usepackage{xspace}
\usepackage{ifthen}

\newcommand{\carlo}[1]
{\ifthenelse{\equal{\showcomments}{true}}
{{\color{blue}{\textbf{Carlo says:} #1}}}{\xspace}}

\newcommand{\andrea}[1]
{\ifthenelse{\equal{\showcomments}{true}}
{{\color{orange}{\textbf{Andrea says:} #1}}}{\xspace}}

\newcommand{\showcomments}{true}

\begin{document}
\title{Time as change}

\author{Marcello {Poletti}}
\email{epomops@gmail.com}
\affiliation{San Giovanni Bianco, Italy}

 \begin{abstract}
According to Aristotle \textit{"time is the number of change with respect to the before and after"}\cite{B1}. That’s certainly a vague concept, but at the same time it’s both simple and satisfying from a philosophical point of view: things do not change along time, but they do change and the measurement of such changes is what we call time. This deprives time of any attribute of substantiality, meanwhile depriving it of all problems in defining the properties of time as a substance.
With the rise of Classical Mechanics, Aristotle's view is abandoned and Newton’s concept of "true"\cite{B2} and absolute time imposes itself; time flows independently on changes of any kind. Relativity will then radically modify our concept of time, but won’t actually modify the fundamental idea: things keep changing along time - changes do not make time.
This work will argue  Aristotle’s thesis, showing how such an approach automatically leads to the principles of Special Relativity.  An interesting consequence and, at least virtually, measurable will also be highlighted: the fact that synchronizing two clocks with a precision greater than a certain scale is impossible, estimating such scale around $10^{-22}s$.

\end{abstract}

\maketitle

\section{Introduction}
The most evident difference between an Aristotle's approach, according to which time is constituted by the totality of changes, and the current approach used in Physics, according to which things change along time, is probably that the previous one doesn’t admit, by definition, a permanent state of rest and the second approach does admit it. 
The first law of motion states that \textit{"a body continues in its state of rest […]"} and this is in clear contradiction of Aristotle’s approach, unless we envisage that the rest state is only, implicitly or explicitly, an approximation for disregarding a series of changes which "maintain active" the time stream.

This is actually compatible with our usual idea of matter. A stone is in a state of rest, according to the current meaning, as its mass center doesn’t clearly move (and the body has not angular momentum) but we imagine it, although at rest, being made of molecules and atoms which aren't actually in a state of rest but on the contrary shake, having a complex internal structure, and constantly exchange electromagnetic radiation by absorption and emission with the surrounding environment. A body in its "state of rest" is actually a continuous succession of events.

Einstein originally writes\cite{B3} his most famous equation opposite to the common use:
\begin{equation*}
	m=E/c^2
\end{equation*}
explicitly stating \textit{"The mass of a body is a measure of its energy content"} as if the mass itself, even if in the state of rest, was thought as containing constant changes.

Penrose\cite{B4} combines Einstein’s equation with the equally famous De Broglie’s equation\cite{B5}, achieving:

\begin{equation*}
	m=h\nu/c^2 
\end{equation*}
and he comments even more explicitly: \textit{"The mass of any stable particle, therefore, determines a very precise clock rate, as given by this frequency."}.

According to this view Newtonian and Aristotelian time can be reconciled, considering the first one an approximation, so that the majority of changes which "measure time" are ignored and changes caused by the motion studied by the mechanics (e.g. a falling stone) have no perceived influence on the total changes which are taking place.

\section{Time as change in Classical Mechanics}
\subsection*{Definition}
The Classical Mechanics typically studies the motion of pointlike particles and, by extension, of spatially extended bodies, and fluids. A typical point particle is an ideal body with no size and a non-zero mass.
In order to organize a Classical Mechanics based on Aristotle’s time we’ll consider Democritus’ massless atoms, in order not to face difficulties previously mentioned in the introduction, atoms in motion inside of a Newton’s Euclidean space.

The only dynamic properties for these point particles are their relative positions, therefore the Aristotle’s definition of time can be easily expressed as:
\begin{equation}\label{main}
	\Delta T=\Delta S/c
\end{equation}
That means, time interval is identified with displacement, up to an arbitrary constant c (c being a velocity) whose only aim here is to lead to a change of unit of measurement.
The previous definition \ref{main} rewritten in differential form can be expressed as:

\begin{equation*}
	dx^2+dy^2+dz^2-cdt^2=0
\end{equation*}
Which corresponds to the metric of light-like paths in a Minkowski space. 

Regarding the speed we notice clearly that the only accepted value is:
\begin{equation*}
v=\frac{\Delta S}{\Delta T}=c
\end{equation*}
\subsection*{Moving Reference Frames}
Let P a path traced by the point, it can be expressed in a parametric form:
\begin{equation*}
	P = \begin{cases}
		x(t) \\
		y(t) \\
		z(t)
	\end{cases}
\end{equation*}
And such form will be called \textit{normal} if
\begin{equation*}
|\dot P|=\sqrt{\dot x^2+\dot y^2+\dot z^2}=c
\end{equation*}
For a path in a normal form we have:
\begin{equation*}
	\Delta S(t)=\mathcal{L}(P,t)=\int_0^t{|\dot P|}=ct
\end{equation*}
And therefore, for a path in a normal form, $t$ parameter corresponds with the given definition of time.

The equivalence space-time defined in \ref{main} makes these two concepts interchangeable. 
Inter alia a Cartesian axis can be identified by the path which travels it, identifying its points with the necessary time for the path to reach them.

Consider then the following normal linear path in a reference frame $k$, with its origin in $O$.
\begin{equation*}
P = \begin{cases}
	ct \\
	0 \\
	0
\end{cases}
\end{equation*}	
Apply therefore a linear transformation of the reference frame $k$ parameterized with the same time $t$ and depending on the parameters $v$ and $\gamma$:
\begin{equation*}
	k' = \begin{cases}
		x'=\gamma (x-vt) \\
		y'=y \\
		z'=z
	\end{cases}
\end{equation*}	
Hence, $P$ reaches the point $(x,0,0)$ at time $t_x=x/c$; we have therefore
\begin{equation*}
	k' = \begin{cases}
		x'=\gamma x\dfrac{c-v}{c} \\
		y'=y \\
		z'=z
	\end{cases}
\end{equation*}	
Applying now the inverse transformation on $k'$, it is required by symmetry that
\begin{equation*}			
	x''=\gamma \left[\gamma x\dfrac{c-v}{c}\right]\dfrac{c+v}{c}=x
\end{equation*}
Hence
\begin{equation*}
	\gamma=\dfrac{1}{\sqrt{1-v^2/c^2}}
\end{equation*}
Finally, time will have a formal transformation in $k'$
\begin{equation*}		
	t'=\dfrac{x'}{c}= \gamma(t-\dfrac{v}{c^2} x)
\end{equation*}
The reference frame $k'$ will be called  \textit{moving with velocity $v$ relative to $k$}.

A moving reference frame follows Lorentz transformations.
\subsection*{Moving paths}
Given now a generic normal path:
\begin{equation*}
	P = \begin{cases}
		x(t) \\
		y(t) \\
		z(t)
	\end{cases}
\end{equation*}					
That can be expressed in an integral form as:
\begin{equation*}
	P = \begin{cases}
		\int\dot x(t)+x_0 \\
		\int\dot y(t)+y_0 \\
		\int\dot z(t)+z_0 \\
	\end{cases}
\end{equation*}				
The infinitesimal shift is:
\begin{equation*}
	dP = \begin{cases}
		\dot xdt \\
		\dot ydt \\
		\dot zdt \\
	\end{cases}
\end{equation*}			
and, if placed within a reference frame $k'$ moving at a $-v$  velocity, it becomes	
\begin{equation*}
	dP' = \begin{cases}
		\gamma (\dot x+v)dt \\
		\dot ydt \\
		\dot zdt \\
	\end{cases}
\end{equation*}						
hence
\begin{equation*}
	P' = \begin{cases}
		\gamma (x+vt-x_0)+x_0 \\
		y \\
		z \\
	\end{cases}
\end{equation*}			
$P'$ path is  equivalent to $P$ within the reference frame $k'$ moving with velocity $-v$, vice versa we will affirm that \textit{$P'$ is equivalent to $P$ moving with speed $v$}.			
\subsection*{Clocks}
$P'$ path obtained in the previous paragraph is a non-normal path, we have
\begin{equation*}
	\dot P' = \begin{cases}
		\gamma (\dot x+v) \\
		\dot y \\
		\dot z \\
	\end{cases}
\end{equation*}								
Hence
\begin{equation*}
	\mathcal{L}(P',t)=\int_0^t{|\dot P'|}=\int_0^t{\sqrt{\gamma^2(\dot x+v)^2+\dot y^2+\dot z^2}}
\end{equation*}
By using the normality of $P$, we have:
\begin{equation*}
	\mathcal{L}(P',t)=\int_0^t{\sqrt{\dfrac{(\dot x+v)^2}{1-v^2/c^2}+c^2-\dot x^2}}
\end{equation*}			
By simplifying
\begin{equation*}
	\mathcal{L}(P',t)=\gamma\int_0^t{c+\dfrac{\dot xv}{c}}=\gamma(ct+v(x-x_0)/c)
\end{equation*}	
Specifically, if $P$ is cyclic, that is if $P(t)=P(t+T)$, we obtain for $P'$ a length of a path of one cycle equal to		
\begin{equation*}
	\mathcal{L}(P',t)=\gamma cT
\end{equation*}													
The time of a $T$ cycle is therefore dilated of a Lorentz factor
\begin{equation*}
	T'=\gamma T
\end{equation*}	
We might consider a cycling path as a clock marking time in relation to its own cycles; according to this, a clock’s mechanism is slowed down of a Lorentz factor.

A clock or a cyclic path can be thought as "objects themselves", choosing to ignore their generating path and studying their overall dynamics.

Such dynamics, as shown above, are subject to the laws of Special Relativity.
\subsection*{Mass}
In order to introduce a mass concept, we consider a circular oscillator of radius $r$, normal, within the plane $(x,y)$:
\begin{equation*}
	C = \begin{cases}
		rcos(ct/r) \\
		rsin(ct/r) \\
		0 \\
	\end{cases}
\end{equation*}
$C$ is an oscillator of radius $r$, pulse $\omega=c/r$ and frequency $\nu=c/2\pi r$.

Using Penrose’s interpretation shown above, we get the definition:
\begin{equation*}
	m=h\nu/c^2 =\hbar/cr
\end{equation*}									
Vice versa
\begin{equation*}
	r=\hbar/cm
\end{equation*}									
In order to get a real world example, if an electron was shaped as a circular path, its radius would be 
\begin{equation*}
	r_e=3.87\cdot 10^{-13}
\end{equation*}								
a value lower than Bohr’s radius $a_0=\hbar/(cm_e \alpha)$ for a factor identical to the inverse of the fine-structure constant, equal to approximately 137.				
\subsection*{Scholium}		
Aristotle's definition of time, applied to Classical Mechanics via the simple definition $\Delta T=\Delta S$ (in natural units), automatically leads to a definition of Moving Reference Frame subject to Lorentz transformations, leading therefore to the definition of moving path and from there to Special Relativity. Note that this happens with no mention of properties of light or of the electromagnetic field. Such a concept leads to two symmetrical observations: on one side, this concept is enough of a prove that defining time as a measure of change is a hypothesis, mainly abandoned by modern Physics, really worth being reconsidered and further explored in different fields of Physics; on the other side, such an approach allows re-reading the Relativity itself under a new light which considers both existence and constancy of a maximum speed not as postulates but as simple consequences of the definition of time itself. Therefore time and space turn out to be deeply interconnected and interchangeable, substantially, by definition.

Through this approach, time loses any substantial characterization (one might say that \textit{"time does not exist"\cite{B6}} as well) and is replaced by the weaker but philosophically more gratifying concept of change.
\section{Multiple paths}
Given two generic paths, $P_1$ e $P_2$, a number of different applications of the concepts of displacement exist; there are therefore various different definitions of the concept of time, as shown thus far.

On one side, $P_1$ and $P_2$ separately define two proper times $t_1$ and $t_2$, as if the other path was negligible; 

on the other side, it is altogether possible to consider the total displacement as time $t_{12}$ of the couple, hence:
\begin{equation*}
	t_{12}=t_1+t_2
\end{equation*}
Clearly, as a consequence of this dissertation’s assumptions, no mechanisms synchronizing $t_1$ and $t_2$ can be accepted. That is, it isn’t acceptable that any absolute clock establishing a defined relationship between the two paths exists. In other words, the idea of "time as change" necessarily implies that knowing $t_1$ is not enough to gather exact information about $t_2$.

Suppose to discretize the frame, postulating that every move (therefore every time) happens along discrete steps. Assigning such steps to any of the two paths randomly seems necessary.

Hence arises the actual problem of computing the probability that when the time of $P_1$ path is $t_1$, then the time of $P_2$ path is $t_2$.

The Classic Probability Theory answers this question with the negative binomial distribution (or Pascal distribution):
\begin{equation*}
	P(t_1,t_2)=\dfrac{\binom{t_1+t_2-1}{t_1-1}}{2^{t_1+t_2}}
\end{equation*}
Hence the probability density that $t_2=xt_1$ can be deduced
\begin{equation*}
	\tilde P(n,x)=\dfrac{\binom{n+nx-1}{n-1}}{2^{n+nx}}=nP(n,nx)
\end{equation*}
Where $n$ is a whole number expressing the size of time quantum $dt=1/n$.

By applying the central limit theorems (De Moivre-Laplace) we get
\begin{equation*}
	\tilde P(n,x)\approx\frac{ne^{-\frac{(n-nx-1)^2}{2(n+nx-1)}}}{\sqrt{2\pi(n+nx-1)}}
\end{equation*}
Given $\tau=1/n$ we have
\begin{equation*}
	\tilde P(n,x)\approx\frac{e^{-\frac{(1-x-\tau)^2}{2(1+x-\tau)}}}{\sqrt{2\pi\tau(1+x-\tau)}}
\end{equation*}
And we can approximate this with a normal distribution with standard deviation
\begin{equation*}
	\sigma=\sqrt{2\tau}
\end{equation*}
If $\tau$ is infinitesimal, so changes happen in continuity, $\sigma$ is infinitesimal as well and times $t_1$ and $t_2$ have a natural tendency to line up. On the contrary, if $\tau$ was subject to a form of quantization, both times would be statistically out of phase of a factor $\sigma$.

In particular, choosing for $\tau$ the value of Planck time ($10^{-44}s$), we obtain
\begin{equation*}
	\sigma\approx10^{-22}s
\end{equation*}
In other words, two clocks couldn’t be synchronizable with a precision greater than $10^{-22}s$ which certainly is an extremely small time value, but not too far from the scale of atomic phenomena.
\subsection*{Scholium} 
If the "measure of change" requested by the Aristotle’s approach is meant as a discrete computing of changes, the limited precision with which two clocks are synchronizable arises. Therefore below such limit the notion of time becomes deeply messy and presumably inapplicable in its traditional meaning.

The usual chance to order events according a before and an after would only be possible from a certain scale up and such scale, as small as it may be, would be much larger than the quantization scale anyway, to the point of being comparable to the shortest time intervals so far measured \cite{B7}. 
\section{Conclusions}
Nature of time is still an important enigma in Natural Philosophy but probably time is "The Enigma" for most physicians and philosophers. I tried to show the great richness that a thousands years old idea might hold. 

This essay is an invitation to try and stop thinking of how things have been changing over time, considering instead how things change and the measurement of such changes is the measurement of time.
This time is structurally relativistic and local, and it has a peculiar extensive nature: the time of two similar objects is approximately twice as much the time of a single object. Everyday time becomes somehow thermic, a sort of an almost constant average of the number of local changes. That’s more than an analogy, both Aristotle’s time and temperature measure change, agitation. 
According to this view, similar to the state of quiet, also the absolute thermic zero cannot be a persistent absolute state as this condition would stop the flowing of time.

\section*{Acknowledgements}
Thanks to Alessandra and Carlo for helping me with the translation of this article.

To Deborah, who challenged time.

\end{document}